# An Open, Multi-Sensor, Dataset of Water Pollution of Ganga Basin and its Application to Understand Impact of Large Religious Gathering

Biplav Srivastava[1,3], Sandeep Sandha[1], Vaskar Raychoudhury[2,4], Sukanya Randhawa[1], Viral Kapoor[2] and Anmol Agrawal[2]

[1]IBM Research – India; [3]IBM T J Watson Research Center, USA
[2]Indian Institute of Technology, Roorkee, India; [4]Universität Mannheim, 68131 Mannheim, Germany

**Abstract**

Water is a crucial pre-requisite for all human activities. Due to growing demand from population and shrinking supply of potable water, there is an urgent need to use computational methods to manage available water intelligently, and especially in developing countries like India where even basic data to track water availability or physical infrastructure to process water are inadequate. In this context, we present a dataset of water pollution containing quantitative and qualitative data from a combination for modalities - real-time sensors, lab results, and estimates from people using mobile apps. The data on our API-accessible cloud platform covers more than 60 locations and consists of both what we have ourselves collected from multiple location following a novel process, and from others (lab-results) which were open but hither-to difficult to access. Further, we discuss an application of released data to understand spatio-temporal pollution impact of a large event with hundreds of millions of people converging on a river during a religious gathering (Ardh Khumbh 2016) spread over months. Such unprecedented details can help authorities manage an ongoing event or plan for future ones. The community can use the data for any application and also contribute new data to the platform.

## Introduction

Water management deals with twin issues of *monitoring and quality control of potable water sources* and *discharge of domestic and industrial effluents*. These two objectives become related when effluents are directly dumped in rivers/lakes and such polluted water becomes the water source for downstream population.

With a growing human population and shrinking supply of potable water, water resources are under intense stress world-wide and especially in developing countries where even basic data to track water availability or physical infrastructure to process water are inadequate. The general public is often aware of the upcoming water crisis and increasingly has access to computing resources, via mobile phones. Consider that *Abhay* may want to take a bath in the river during a religious festival and would want to know which banks (religious sites, i.e., *ghats*) of the river are feasible to go without getting sick. *Bina* may want to tap ground water or fetch from river water for household activities. *Chetan* may want to use river water for irrigating his fields over ground water. *Divya* may wonder if fishing or vegetable growing is promising on the river catchment area to supplement her family's earnings. These and other users are routinely taking decisions which can be driven by water pollution data if it were available and AI-researchers could create decision-support apps.

The situation gets further aggravated when large-scale gatherings take place on the banks of rivers and people indiscriminately use water in activities bathing and religious rites where they may discharge external additives to water like coins, milk, flowers and organically ashes. Ironically, water data has been collected by numerous agencies around the world (usually government, academic and non-profits) and individuals for many years on different stretches of the river measuring different parameters at different times using different technologies. However, they as a whole are not only insufficient in terms of quality and quantity but also inaccessible to wider public because they were not annotated along common usage scenarios. Water data is available online from a few regions of countries like Bath in UK (UK 2016), New South Wales in Australia (Australia 2016) and Hudson Valley in USA (USA 2016). In India, CPCB maintains a website with data from 10s of locations since 2015.

In this context, our contributions in paper are that we:

- Release and present a dataset of water pollution from Ganga basin in India that contains both quantitative and qualitative estimates obtained from a combination for modalities - real-time sensors, lab results, and estimates from people using mobile apps. It has thousands of datapoints along with permissible parameter limits and their applicability for different purposes which can be ac-cessed digitally via APIs, mobile apps as well as updated by others (urls omitted during blind review phase).
- Demonstrate usage of released data in a case study where hundred of millions of people assemble along a river for over 4 months for religious rituals and understand its environmental impact.
- Discuss methodological and AI issues in using the released data for other use-cases.

In the rest of the paper, we discuss the dataset and how it was created, followed by demonstration of its usage in understanding impact of a large event and concluding with a discussion on issues in using the data for other use-cases.

## Data Set

In this section, we describe the quantitative and qualitative datasets released publicly by us and the methodology followed. We start with a background of water pollutants and how they are measured. We then describe the released data and its characteristics.

**Water Quality Measurement Parameters**
When monitoring water, a number of parameters are of interest to different stakeholders depending on their purpose. Environment agencies like EPA in US and CPCB in India recommend tens of parameters - CPCB recommends more than 30 parameters (CPCB 2014a, CPCB 2014b) in India. We describe the most important here - *pH*, *electrical conductivity*, *dissolved oxygen*, and *turbidity*. Agencies also recommend standards for different usage like drinking, irrigation - CPCB prescribes parameters of interest and their ranges for 22 industries (CPCB 2014b). A major challenge is overlapping specifications of multiple agencies within a government and also at multiple levels (national, state and international) which can be in conflict. We reconcile ranges for 25 parameters in the data released using CPCB guidelines on purpose and pollutants.

- *pH:* pH is a measure of how acidic/basic water is. The range goes from 0-14, with 7 being neutral. The pH scale is a logarithmic one; so, water with a pH of 5 is ten times more acidic than water having a pH of 6.
- *Electrical Conductivity:* Pure water is not a good conductor of electricity. Conductivity of water increases as the concentration of ions increases. While drinking water has a conductivity of 0.005 – 0.05 S/m, for sea water it is 5 S/m. Conductivity of water increases with increase in temperature and vice versa.
- *Dissolved Oxygen (DO):* It refers to the level of free, non-compound oxygen present in water or other liquids. Adequate DO is required to maintain good water quality. DO may increase with rising water temperature or decrease by mixing untreated or partially treated sewage with water. As DO level in water drops below 5.0 mg/l, aquatic life is put under stress.
- *Turbidity:* It measures relative clarity of a liquid by the amount of light that is scattered by impurities (like, clay, silt, finely divided inorganic and organic matter, algae, soluble colored organic compounds, and plankton) suspended in the water. Higher level of turbidity in drinking water increases the chance of gastro-intestinal diseases.

*Table I Water standards for different Activities*

| Activity | pH | Conductivity (μS/cm) | DO (mg/L) | Turbidity (FNU) |
|---|---|---|---|---|
| Drinking | 6.5-8.5 | ≤ 2250 | ≥ 6 | 1-5 |
| Bathing | 6.5-8.5 | -- | ≥ 5 | -- |
| Irrigation | 6.0-8.5 | ≤ 2250 | ≥ 6 | 1-5 |

Methods of Collecting Water Pollution Data

- *Sample collection and lab tests for quantitative results:* The most accurate way of measuring water pollution is by collecting water samples from water bodies following standardized practices periodically, testing them for all parameters important for a purpose and publishing them. Such data may be available in printed form but not accessible digitally, is quite expensive to create due to cost of sample collection and lab testing, is time-consuming and quantitative. The scientific challenges are in testing the pollutants and collection of samples. We took published lab results from over 60 locations in India from different sources like universities and put them on our common platform. They covered 4-5 parameters at a time and the samples were collected at frequencies varying from once a month to yearly.
- *Real-time (RT) sensing:* We have used a state-of-art, portable, commercial multi-sensor kit from Hanna (HI9829[1]) which monitors up to 14 different water quality parameters including pH, ORP, conductivity, dissolved oxygen (DO), turbidity, ammonium, chloride, nitrate, and temperature. It is possible to log the data locally and to track location of measurement points using GPS (up to an accuracy of 10m). Following convention, the sensor kit is pre-calibrated to report the readings at 25C to remove the effect of temperature on the readings captured on different days and times. Note that RT sensors can only measure a subset of supported parameters (typically 4-7 based on probe deployed) at a time and the technology does not support measurement of all parameters an agency may recommend like Biological Oxidation Demand (BOD) and Faecal Count (FC). RT sensors, once procured, are inexpensive to operate and give results instantly. However, the technical challenge with them is determine a reliable process to measure under an environmental condition. We used our sensors on Ganga and its tributary, Yamuna, in India using a non-stationary platform to get higher spatial coverage, in contrast to the normal practice of attaching sensors to a fixed location. Further, we used a mobile app (explained below) to capture qualitative data in conjunction with the RT sensors at select landmarks to help with validation.

[1] ¹ http://hannainst.com/

• Crowd-sourced, qualitative, sensing using a mobile application: We have released a mobile application (app) on Google play store called *Neer Bandhu* (NB), meaning 'Water Friend', which can be used by people to take the picture of a river area depicting water pollution (Fig 1-left) and to select a few options regarding the characteristic of water (Fig 1-middle) which are specially derived from standards and common practices that exist in the water domain. The data that is being collected through this app is available within the app (Fig1-right) and online. It has to be noted that the qualitative data is not precise as the quantitative data. However, it offers evidence that can be used to interpret results from other modalities like RT sensors and also capture information that people care about but are hard to measure (e.g., water smell).

We used *Neer Bandhu* both independently as well as in conjunction with RT sensors.

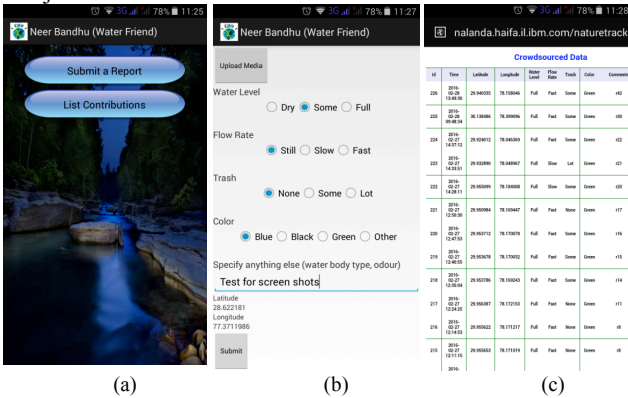

(a)　　　　　(b)　　　　　(c)

*Fig. 1   Qualitative Data Collection using Neer Bandhu*

**Data Released and Exploration Tools**

Figure 2 shows the location of quantitative and qualitative data on a map. We have 117 data points from lab tests, 10,506 from RT sensors and 78 points from NB. It covers the Ganga basin covering Ganga and its tributary, Yamuna. All data, except lab results, was collected by authors from Ganga (Feb-April 2016), Yamuna (Dec 2015) and Hindon, a tributary of Yamuna (Sep 2015). We commissioned and release 5 of our lab results and the rest were contributed by other teams.

The data is accessible via an Application Programmable Interface (API) to our cloud platform called BlueWater so that a consuming application can get all changes seamlessly and automatically. We have also released a mobile app, called GangaWatch, which uses the public APIs, to all exploration of released data.

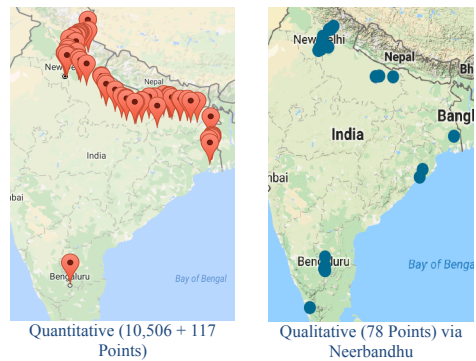

Quantitative (10,506 + 117 Points)　　Qualitative (78 Points) via Neerbandhu

*Fig. 2   Statistics and locations of pollution data*

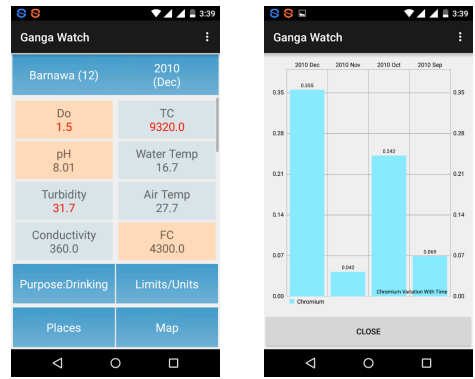

*Fig. 3 GangaWatch screenshots showing released data*

## Application Case Study

In this section, we discuss our analysis of Ganga water quality during Ardh Kumbh 2016 held at Haridwar. We also describe the relevant data collection process in more detail to help interpretation of the results.

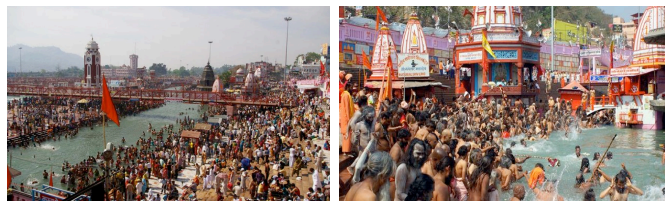

*Fig. 4   Religious Gathering during Ardh Kumbh 2016, Haridwar*

## Background of Khumbh – A Large-Scale Religious Event near a River Bank

In India, there are several river-centric large-scale religious gatherings of which the Kumbh festivals are most important. Here, millions of people gather at the nearest holy places and undergo ablution in the river on specific days and times with religious motifs. People indiscriminately use water for bathing and religious activities, such as, offering milk, ghee, flowers, coins, idols, ashes of departed ones, body hairs into water (Kulshrestha and Sharma

2006). Depending on their activities, the river water gets affected impacting economic activities downstream.

Ardh Kumbh 2016 took place in Ganga river in Haridwar, Uttarakhand between Jan-April. An estimated 10 crore (100 Million) devotees took part overall with nine peak bathing days when 10s of millions devotees came[2].

**Data Collection**

The study involved collection of water quality parameters at a high resolution as detailed in Table II over 76 Km length along the Ganga river and its canal in Rishikesh, Haridwar and Roorkee regions of the Uttarakhand district in India. We selectively chose multiple locations and monitored them throughout the festival. The choice of location was made such that water quality *before*, *after* and *at* the bathing places could be studied.

*Table II      Data Colletion Location Statistics*

| Total no. of collection coordinates (repeating) | 134 (only on the Ganga and its Canal) |
|---|---|
| Starting coordinate | (30.13848, 78.39982) – Near Shivpuri, Rishikesh |
| Last coordinate | (29.81985, 77.87248) – Asaf Nagar, Roorkee |
| Total Road distance | 76.1 KMs |

We collected data on the following days (Table III) in February to April 2016.

*Table III    Data collection dates*

| Date | Kumbh Bathing Day/ Other day | Place | Time |
|---|---|---|---|
| Feb. 27 | Other day | Haridwar and Roorkee | 3-7 PM |
| Feb. 28 | Other day | Shivpuri, Rishikesh, Roorkee | 11 AM - 6 PM |
| Mar. 07 | Bathing Day | Haridwar | 6-10 AM |
| Apr. 07 | Bathing Day | Haridwar | 1-4 PM |
| Apr. 15 | Bathing Day | Haridwar | 7-10 AM |

In most of the places, we took readings by standing on the bank but in Rishikesh we used boats for data measurements in the river. We chose both peak and off-peak bathing days to see the variations caused by mass bathing. On peak bathing day(s), data was only collected at Haridwar since it is the center of almost all religious activities.

**Analysis and insights**

In this section, first we show how the water quality parameters vary for the entire stretch from Rishikesh (Shivpuri) to Roorkee and then we shall show how the parameters vary during the bathing days at Har-ki-Pauri location, Haridwar, which is a prominent location midway.

**Variations over the Entire Stretch**

In order to understand the effect of bathing at Haridwar, we followed the river Ganga upstream and downstream. In Fig 5, the mean of parameter values at different places is plotted. Haridwar stands out distinct from the upstream Rishikesh and downstream Roorkee. Ganga at Rishikesh (Shivpuri) is still at the Himalayas and has rocky bed and high velocity.

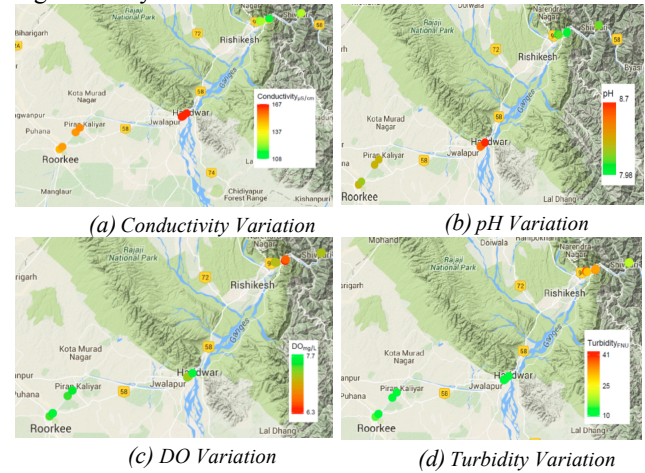

*(a) Conductivity Variation*      *(b) pH Variation*

*(c) DO Variation*      *(d) Turbidity Variation*

*Fig. 5    Heat Maps of Parameter Variation (over entire stretch)*

So, it carries large amount of silt, which affects its *turbidity* (Fig. 5(d)). At Haridwar, extensive use of water is affecting its *conductivity* and *pH*, as seen in the plots (Fig. 5(a) and (b)). It can be noted that, over the entire stretch the *DO* in Ganga was much within the safe limits, although it shows the variations across different *ghats* (religious site), places and banks (Fig. 5(c)).

**Variation Only at the Bathing Ghats in Haridwar**

The variations of different parameters are shown in Fig. 6 using heat maps. From top to bottom, each column has six heat maps showing day-wise variation of pH, conductivity, DO and turbidity, followed by Water Quality Index (WQI), and cluster of data points into good (Cluster 1) and poor (Cluster 2) quality ones. From the plots it can be seen that different bathing places at Haridwar are utilized to different extent. Minor variations along the stretch are clearly visible. As seen in the plots water quality on both banks of the rivers also shows significant variation. This variation is replicated on each bathing day, which clearly shows that one bank of the river is utilized more whereas other bank is utilized less. This is also expected because one bank of the river has Har-ki-Pauri (a place most sought after for religious baths) and other ghats, where there is less activity on the other bank of the river. Multiple readings (around 30) were collected at each place and mean of the values was used to create the plot. Different days also shows variations which indicates that each day variable amount of people attend the religious bathing. Our plots show changes which can be reasoned after understanding the effect of bathing as well as the river geological settings.

**Water Quality Analysis during Ardh Kumbh 2016**

Water Quality Index (WQI) defines the overall water quality calculated using several ways depicted in (Tyagi 2013). In order to compare the water quality at different places during the study on Ganges, we have calculated the **Weighted Arithmetic Water Quality Index** as shown in (Tyagi 2013). The calculation of WQI is influenced by the selection of limits of water quality parameters and their ideal values. The ideal value and limits of water quality parameters used in WQI calculation is as per Table I.

---
[2] http://www.kumbhamela.net/kumbha-mela-haridwar.html

TABLE IV  THE INTERPRETATION OF WQI (TYAGI 2013)

| WQI Value | Rating of Water Quality | Grading |
|---|---|---|
| 0-25 | Excellent | A |
| 26-50 | good | B |
| 51-75 | poor | C |
| 76-100 | Very poor | D |
| Above 100 | Unsuitable for Drinking | E |

The WQI plots (Fig. 6(e)) depicts that relatively poor water quality is observed around Har-ki-pauri and rather cleaner water quality is available before and after Har-ki-Pauri. The WQI plots show that on different day(s), water quality varies at different Ghats. The variations are clearly highlighted by the heat map of the plots. These finer variations are captured by collecting water quality every second over these places then taking mean of the values.

In order to understand the variation of WQI at different places we used K-means clustering. So as to differentiate between regions having good water quality and poor water quality we computed two centers as shown in plots (Fig. 6 (f). The cluster plots show the places having similar WQI, in same color. The variations are as expected, with the sites which are used less frequently are clubbed together whereas the sites which are used more shows high WQI. The variations on different days are different. Also on each day, WQI varies between different sites as shown by the plots.

## Discussion and Conclusion

As one uses released data, one needs to overcome a few challenges: (a) *Overlapping availability of usable pollution data by pollutants, data type and locations*. The data is large by relative volume but the sample size, frequency and measurement process vary over time and may even be under dispute when judged later as technology evolves. Many historical readings are once per month or quarter and the process used may vary. There is a wide diversity in the number of parameters recommended to be measured and the subset actually tested. Our stand is to consider all available data and correlate them (via statistics and machine learning) to establish data quality. Further, mobile allows capture of qualitative perception of pollution from people which can help validate operation of sensors. This needs combination of techniques that can handle quantitative and qualitative data, respectively. (b) *Ambiguity of purpose in using water pollution data*. The data publishers do not explicitly and consistently clarify what purpose, like drinking or distillery industries, their data is suitable to be used for. The consuming application should use all annotations where applicable, but should also consider raw sensor data when annotation is missing and label it. (c) Ambiguity of permissible limits based on a purpose since there may be overlapping standards at local, national and international levels. We have resolved published data by CPCB standard but other standards may be relevant for some users. (d) There was no easy access to pollution data via mobile apps and APIs, and GangaWatch and BlueWater fill the gap, respectively. (e) Reconciling multiple parameters given sparsity of coverage to get aggregate view. In the Khumbh case, we could use WQI to get aggregate view since all parameters of interest were present.

In related work, human activities on rivers in general, and Kumbh festivals on Ganga in particular, have been regularly studied in the past (Karar 2010, Khanna, et al 2010, Sharma, et al 2012, Arora 2010, Singh and Bisht 2013, Matta 2014, Panwar et al 2015, Bhutiani 2016). However, such studies were restricted to taking water samples at select locations and analyzing them in labs for local analysis. Our released data can be used for generating new insights like detailed spatio-temporal outlook of the river or projections at a place over time (as we demonstrated), validating data quality, manage an ongoing event or plan for future ones.

More importantly, other researchers can use available data, both quantitative and qualitative, to create insights and decision aids in more use-cases. Further, they can themselves collect new data and contribute back via APIs to the platform so that it is available to the community.

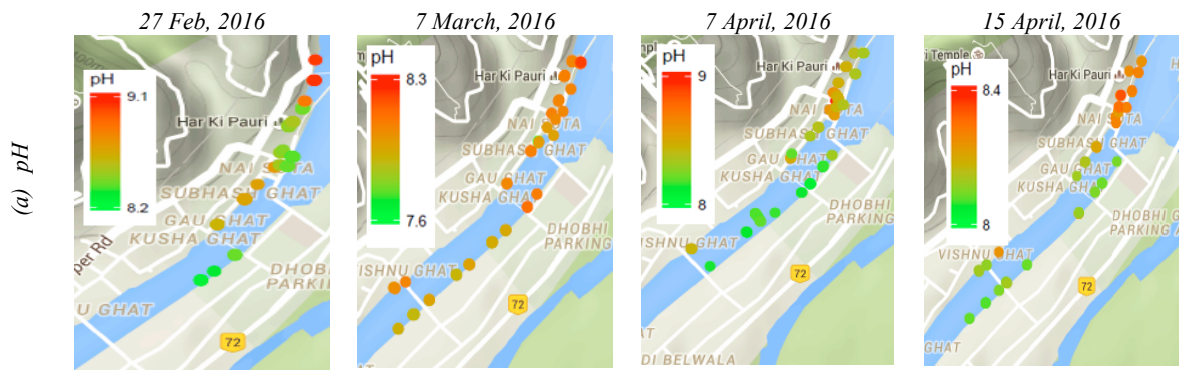

*27 Feb, 2016*   *7 March, 2016*   *7 April, 2016*   *15 April, 2016*

*(a) pH*

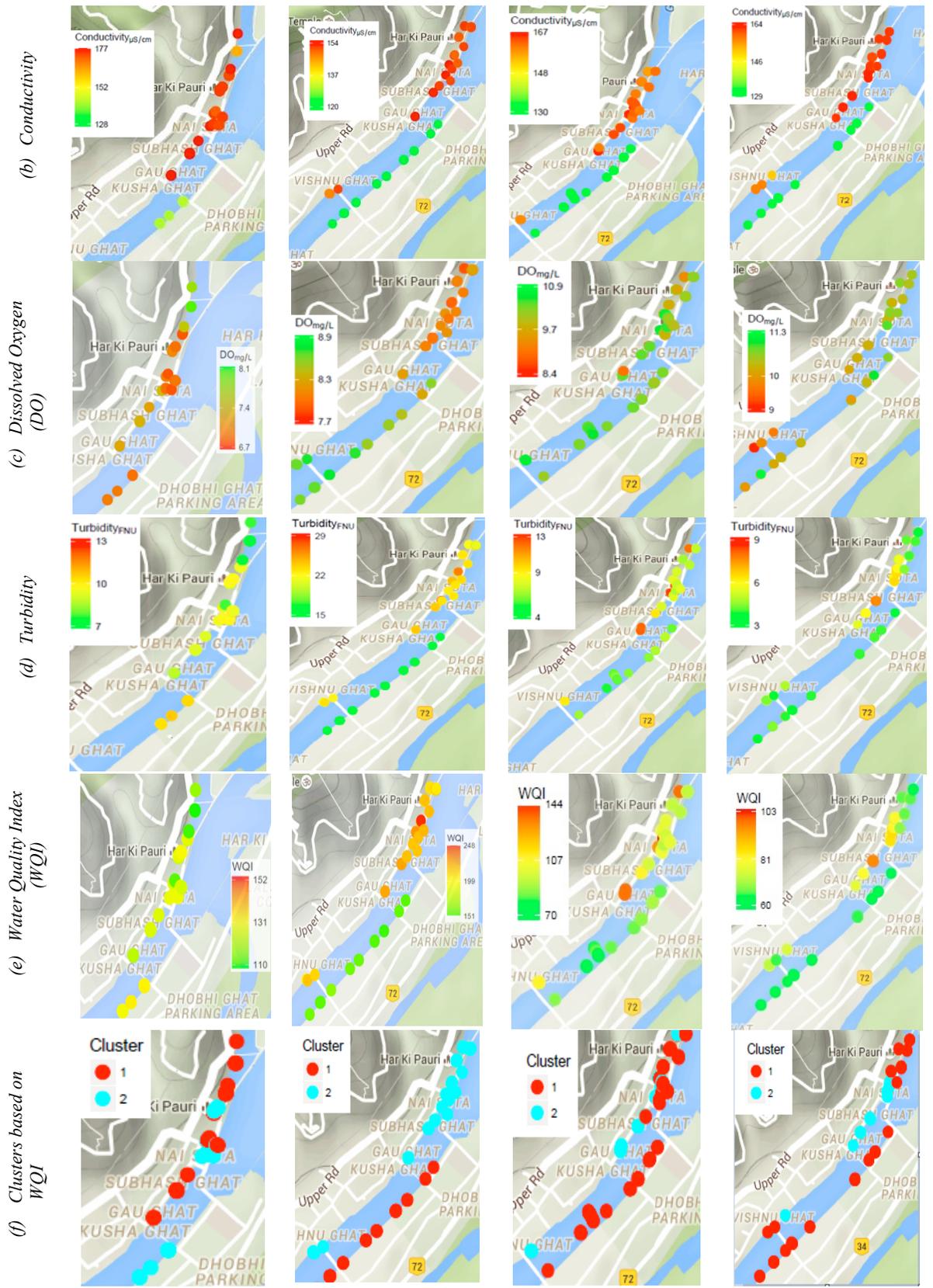

Fig. 6 Heat Maps to Show Day-wise variation of pH, Conductivity, DO, Turbidity, WQI, and Clusters during Ardh Kumbh 2016